\documentclass[sigconf, nonacm]{acmart}  

\usepackage{natbib} 
\usepackage{wrapfig} 
\usepackage{enumitem} 

\usepackage{listings}

\setlist{nolistsep}
\AtBeginDocument{%
  }

\copyrightyear{2025}
\acmYear{2025}
\setcopyright{cc}
\setcctype{by}
\acmConference[IUI '25]{30th International Conference on Intelligent User Interfaces}{March 24--27, 2025}{Cagliari, Italy}
\acmBooktitle{30th International Conference on Intelligent User Interfaces (IUI '25), March 24--27, 2025, Cagliari, Italy}\acmDOI{10.1145/3708359.3712072}
\acmISBN{979-8-4007-1306-4/25/03}

\newcommand{\tool}{\textsc{HEPHA}}

\newcommand{\revise}[1]{\textcolor{black}{#1}}
\begin{document}

\title{HEPHA: A Mixed-Initiative Image Labeling Tool for Specialized Domains}

\author{Shiyuan Zhou}
\authornote{These authors contributed equally to this research.}
\email{szhou20@uci.edu}
\orcid{0000-0002-2293-1717}
\affiliation{%
  \institution{University of California, Irvine}
  \city{Irvine}
  \state{California}
  \country{USA}
}

\author{Bingxuan Li}
\authornotemark[1]
\email{bingxuan@ucla.edu}
\orcid{0000-0002-9193-5308}
\affiliation{%
  \institution{University of California, Los Angeles}
  \city{Los Angeles}
  \state{California}
  \country{USA}
}

\author{Xiyuan Chen}
\authornotemark[1]
\email{chen4851@purdue.edu}
\orcid{0000-0003-1315-1978}
\affiliation{%
  \institution{Purdue University}
  \city{West Lafayette}
  \state{Indiana}
  \country{USA}
}

\author{Zhi Tu}
\email{tu85@purdue.edu}
\orcid{0000-0001-6175-164X}
\affiliation{%
  \institution{Purdue University}
  \city{West Lafayette}
  \state{Indiana}
  \country{USA}
}

\author{Yifeng Wang}
\email{yifengw3@andrew.cmu.edu}
\orcid{0000-0002-2559-9908}
\affiliation{%
  \institution{Carnegie Mellon University}
  \city{Pittsburgh}
  \state{Pennsylvania}
  \country{USA}
}

\author{Yiwen Xiang}
\email{yiwenxia@andrew.cmu.edu	}
\orcid{0000-0002-3906-1391}
\affiliation{%
  \institution{Carnegie Mellon University}
  \city{Pittsburgh}
  \state{Pennsylvania}
  \country{USA}
}

\author{Tianyi Zhang}
\email{tianyi@purdue.edu}
\orcid{0000-0002-5468-9347}
\affiliation{%
  \institution{Purdue University}
  \city{West Lafayette}
  \state{Indiana}
  \country{USA}
}

\begin{abstract}
Image labeling is an important task for training computer vision models. In specialized domains, such as healthcare, it is expensive and challenging to recruit specialists for image labeling. We propose HEPHA, a mixed-initiative image labeling tool that elicits human expertise via inductive logic learning to infer and refine labeling rules. Each rule comprises visual predicates that describe the image. HEPHA enables users to iteratively refine the rules by either direct manipulation through a visual programming interface or by labeling more images. To facilitate rule refinement, HEPHA recommends which rule to edit and which predicate to update. For users unfamiliar with visual programming, HEPHA suggests diverse and informative images to users for further labeling. We conducted a within-subjects user study with 16 participants and compared HEPHA with a variant of HEPHA and a deep learning-based approach. We found that HEPHA outperforms the two baselines in both specialized-domain and general-domain image labeling tasks. Our code is available at https://github.com/Neural-Symbolic-Image-Labeling/NSILWeb.

\end{abstract}



\begin{teaserfigure}
  \includegraphics[width=\textwidth]{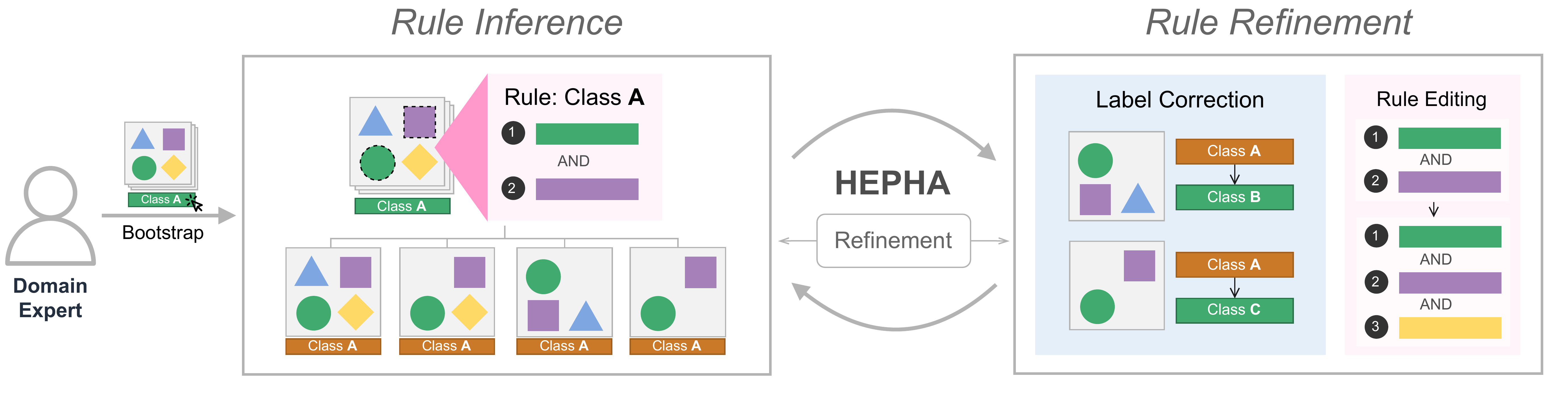}
  \caption{{\tool} is a mixed-initiative tool that supports image labeling by eliciting labeling knowledge from domain experts. Starting with a small set of labeled images, {\tool} generates labeling rules and applies them to all unlabeled images during the \textit{Rule Inference} stage. \revise{Users can then apply their domain expertise to iteratively correct labels or directly edit the inferred rules to improve image labeling accuracy during the \textit{Rule Refinement} stage.}}
  \Description{HEPHA}
  \label{fig:teaser}
\end{teaserfigure}

\maketitle

\section{Introduction}
Deep learning models in computer vision require a large number of labeled images for training. While crowdsourcing is a commonly adopted method for image labeling, it is challenging to find and recruit crowd workers for specialized tasks that require domain expertise. To address this challenge, prior work has proposed to train automated image labeling models to reduce manual effort. \cite{Gu_Leroy} However, these methods still require domain experts to label many images to train the initial models \citep{Ran_2020,Thompson_2020}. For example, \citet{Asaok_2019} proposed to leverage CNN and transfer learning to automatically label early glaucoma eye images. This model requires 4316 eye images for pre-training and 178 eye images for fine-tuning. Furthermore, when the model makes a labeling error, domain experts have a difficult time diagnosing and refining the model due to the lack of transparency of DL models \cite{Ran_2020}.

To address these limitations, we propose a mixed-initiative image labeling tool called {\tool}. {\tool} elicits the labeling knowledge from domain experts as first-order logic rules via inductive logic programming (ILP)~\cite{Muggleton_1994}. Each labeling rule is composed of visual predicates that describe the image, such as objects detected in the image. {\tool} automatically applies these rules to new images to infer their labels. Compared with DL-based labeling methods, {\tool} has two unique advantages. First, ILP only requires a small number of instances, such as 5 images, to infer a logic rule, which makes it much easier to bootstrap the labeling process. Second, the labeling rules are inherently interpretable. Users can easily figure out why {\tool} infers a certain label for an image, diagnose a labeling error, and make corrections. 

Furthermore, {\tool} allows users to iteratively refine the rules by directly editing the rules or providing more labeled images. To facilitate rule refinement,  {\tool} visualizes labeling rules in a visual programming interface and allows users to edit rules through simple drag-and-drop operations. Furthermore, {\tool} provides users with three kinds of recommendations. First, {\tool} recommends which rule needs more refinement based on its accuracy on the holdout dataset. Second, {\tool} uses a saliency-based method to recommend which object should be considered when refining a rule predicate. Third, for users who find it difficult to edit rules directly, {\tool} uses a multi-criteria active learning method to recommend informative and diverse images for user labeling to regenerate labeling rules. 

To investigate the usefulness and efficiency of {\tool}, we conducted a within-subjects user study with 16 participants. While {\tool} is designed for labeling images from highly specialized domains, we are also interested in how well it performs in general domains. Therefore, in this user study, we included four labeling tasks: two from highly specialized domains and two from general domains. We compared {\tool} with the ResNet model \cite{He_Zhang_Ren_Sun_2016} and one variant of {\tool}, which disabled rule editing features. According to the user study results, {\tool} achieved an average labeling accuracy of 87.43\% to 91.42\% on both specialized and general domain tasks while requiring 13 to 17 minutes for labeling. In particular, {\tool} significantly outperformed the deep learning approach in terms of both accuracy and time across all labeling tasks. However, we found that {\tool} required 6 minutes more labeling time than its variant, which may be due to more interactions during rule editing. These findings confirm that {\tool} is a useful and efficient image labeling tool. Refined by human expertise, inductive logic labeling rules can effectively enhance the image labeling process.

Overall, our work makes the following contributions:
\begin{itemize}
    \item We developed {\tool}, a mixed-initiative interaction system that enables users to label images by interpreting the labeling process and refining labeling rules according to human expertise.
    \item We investigated three recommendation features to assist and guide users during the image labeling and rule refinement process.
    \item We did a within-subjects user study to evaluate the usefulness and efficiency of different features in {\tool}.
\end{itemize}

\section{Related Work}

\subsection{Image Labeling}
Crowdsourcing is widely used for image labeling tasks \cite{welinder2010online, sorokin2008utility, welinder2010multidimensional, su2012crowdsourcing}. However, general platforms like Amazon Mechanical Turk often lack domain-specific expertise. Specialized platforms have emerged to address this limitation \cite{meyer2016crowdsourcing, reinecke2015labinthewild}. For example, CrowdMed facilitates crowdsourcing medical diagnoses for patients with undiagnosed illnesses \cite{meyer2016crowdsourcing}, and LabintheWild enables recruitment of participants with specific expertise from diverse backgrounds \cite{reinecke2015labinthewild}. However, these approaches are expensive, requiring the recruitment of costly experts to ensure accurate and reliable labeling. 

With the advent of deep learning, various methods have been developed to reduce human effort \cite{9318911}. These methods can be categorized into {\em fully automated methods} and {\em semi-automated methods}. First, fully automated methods rely entirely on algorithms for labeling \citep{ kalayeh2014nmf, 10.1145/3544548.3581185, marini2024automatic, 9191320}. For example, \citet{kalayeh2014nmf} combines Non-negative Matrix Factorization and K-Nearest Neighbors to achieve automatic image labeling. However, these approaches still require a large volume of images to train labeling models, which can be challenging to collect for domain-specific tasks. Second, semi-automated methods integrate human feedback to enhance labeling accuracy \cite{10.1145/3473856.3473993, Dutta2018AutomaticIA, 10.1145/3511176.3511207, 10.1145/3658271.3658329}. For instance, \citet{Dutta2018AutomaticIA} proposes an interactive system that proposes initial automated labels and facilitates iterative refinement by annotators to achieve higher accuracy. However, previous approaches have two critical limitations: (1) Primitive refinement support: Existing methods often lack innovative mechanisms for efficient label refinement. Current tools propose automated labels but provide limited support for subsequent human refinement, impeding the overall labeling efficiency. (2) Limited interpretability: For domain-specific data, human expert input is invaluable. However, the lack of transparency and interpretability in algorithmic decision-making processes impedes effective domain knowledge elicitation. For instance, \citet{10.1145/3511176.3511207} trained models using expert-annotated data for automated labeling, but the models' complexity can make error correction difficult and limit the use of human expertise.

\revise{To address these limitations, we design an interactive labeling system tailored for specialized domains. 
Our system uses the ILP-based image classification algorithm from RAPID \cite{Wang_Tu_Xiang_Zhou_Chen_Li_Zhang_2023}. However, RAPID lacks a user-friendly interface, which poses challenges for domain experts in effectively using the system.
Therefore, we design a user interface to visualize labeling rules and enable experts to understand and edit the rules. 
We further propose a rule-level recommendation mechanism to assist in rule refinement.
Compared to RAPID, our framework not only improves image labeling accuracy but also enhances interpretability, enabling domain experts to better understand and correct errors.}

\subsection{Interactive Support for Image Labeling}
Recent efforts have introduced interactive tools to facilitate collaboration between automated labeling systems and human annotators \cite{10.1145/3655755.3655760, zhang2023peanut, 10.1007/978-3-031-21707-4_26, gygli2020efficient, 10.1145/1821748.1821847, li2024labelaid}. These tools provide interactive support in the following aspects.

One line of research aims to provide interactive support for the iterative refinement process \cite{10.1145/3655755.3655760, 10.1145/3544548.3581185, 10.1145/3025171.3025208, 10.1007/978-3-031-21707-4_26}. Current works offer such support in two primary ways: First, existing works recommend images for labeling with confident-based approaches or active learning \cite{lynnette2020cross, diaz2024monai, 10.1007/978-3-031-21707-4_26}. For instance, \citet{10.1007/978-3-031-21707-4_26} provides the system confidence level to help users quickly and correctly identify the images that need to be inspected. Second, existing approaches use data visualization to assist users in refining annotations more effectively. For example, \citet{10.1145/3655755.3655760} introduced EasyClick, an interactive tool that provides real-time feedback via a visualization dashboard, enabling annotators to quickly identify and correct errors.

Another line of research aims to optimize annotation workflow. Previous works mainly contribute in two areas. First, multi-modal interaction \cite{gygli2020efficient, 10.1145/3343031.3350535} allows systems to process inputs from various channels, such as speech, gestures, and visual cues, creating a more intuitive user experience. \citet{gygli2020efficient} developed a system that enables users to annotate images by pointing and speaking, integrating gestures and voice commands to streamline the workflow. Second, auto-completion of labels \cite{wong2015smartannotator, dias2019freelabel} simplifies labeling tasks by reducing manual input. For example, \citet{dias2019freelabel} introduced FreeLabel, a tool that converts users' freehand scribbles into precise labels, thus improving both efficiency and accuracy.

In addition to the aforementioned methods, there are also studies focusing on enabling multiple users to collaborate on the same project simultaneously, thereby enhancing labeling efficiency. For example, \citet{10.1145/1821748.1821847} designed an online platform to improve the multi-user labeling process with in-system communication functions, allowing real-time feedback from other users.

Compared to existing approaches, our work introduces a new perspective to elicit the domain knowledge as a labeling rule, which allows us to design more interactive support. For example, {\tool} provides intelligent recommendations to assist users in rule modification and label correction, enabling rapid refinement. Additionally, we have developed a visual programming component to facilitate rule editing. This integrated approach not only enhances user control and efficiency but also improves the interpretability of the annotation process as an additional benefit.

\subsection{Human-AI Interaction}
The interactive support system for image labeling exemplifies human-AI interaction, as it combines the efficiency of AI algorithms with the expertise of human users. Many studies have explored ways to enhance human-AI interaction~\cite{amershi2019guidelines, wang2019human, liao2020questioning, dudley2018review, cai2019hello, cai2019human}. For instance, \citet{amershi2019guidelines} derived 18 design guidelines for human-AI interaction based on over 150 AI-related design recommendations from both academia and industry. The design of {\tool} follows these recent guidelines and principles for human-AI collaboration, with a particular focus on interpretability and user control.

In recent years, significant advancements have been made in enhancing the interpretability of AI models. For example, Gamut is a visual analytics system designed to help data scientists interpret machine learning models \cite{10.1145/3290605.3300809}. \citet{cai2019effects} explores how different types of example-based explanations can influence user perceptions of machine learning systems. Compared with existing approaches, {\tool} enhances interpretability through two new aspects. First, the incorporation of Inductive Logic Learning elicits user knowledge into the automated labeling process, ensuring that the system effectively leverages human expertise for improved accuracy and reliability. Second, by displaying the effects of rules in real time, the rule visualization feature aids users in understanding the impact of each rule on the annotation process. These innovative features align with established guidelines emphasizing clarity in system behavior. 

Additionally, user control in AI-assisted systems has been enhanced through various innovative approaches. \citet{10.1145/2678025.2701399} explored direct manipulation techniques, which allow users to adjust feature importance and model parameters, thereby increasing the controllability of the model. \citet{10.1145/1753326.1753529} further extended this concept by enabling users to interact with confusion matrices, influencing the training objective for real-time optimization based on user-defined goals. Furthermore, tools like PhotoScout \cite{barnaby2024photoscout} integrate a neuro-symbolic domain-specific language (DSL), bridging the gap between user inputs and the underlying search engine to enable complex query execution. Following a similar pattern, {\tool} empowers users with greater control by two strategies: (1) Direct Rule Manipulation: The system enables users to directly modify rules, clauses, and predicates, providing full control over the annotation process. (2) Intelligent Recommendations: The system provides targeted recommendations for uncertain images, rule refinements, and important predicates, facilitating informed decision-making by users. These strategies align with guidelines for efficient correction, simplifying the process of editing, refining, or recovering from AI system errors. Consequently, they reduce user workload and enhance overall accuracy, striking a balance between automation and user agency in the annotation process.

\begin{figure*}[htbp]
  \centering
  \includegraphics[width=0.9\textwidth]{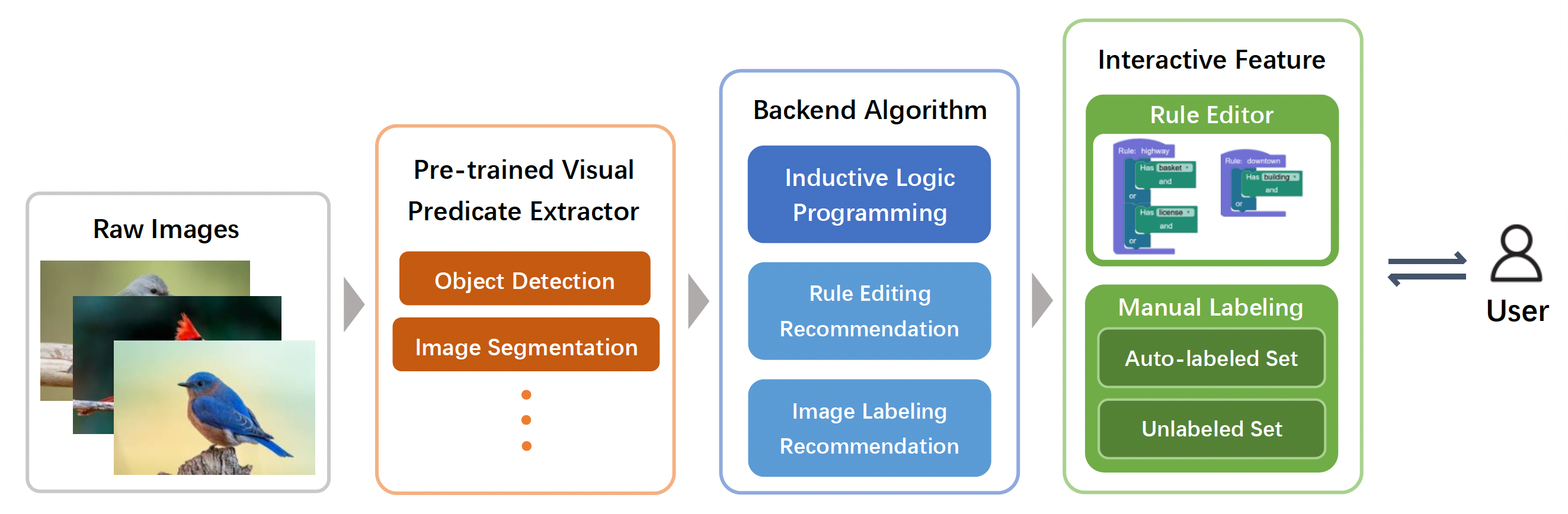}
  \caption{\revise{Overview of {\tool} system architecture.}}
  \Description{System Architecture}
  \label{fig:sys_arc}
\end{figure*}

\section{Design Goals}
In developing {\tool}, we identified four key design objectives. These were informed by our analysis of existing image labeling methodologies \cite{10.1145/3655755.3655760, 10.1007/978-3-031-21707-4_26, welinder2010multidimensional}, prior research on image labeling tasks \cite{su2012crowdsourcing, askhtorab_aiassisted_2021, ratner2017snorkel}, and our empirical experience in the field.

\textit{DG1. \textbf{ Optimizing Efficiency with Limited Data}} Current deep learning methods often require large amounts of labeled data to achieve high accuracy \cite{kalayeh2014nmf, 10.1145/3544548.3581185, marini2024automatic, 9191320}. However, in specialized domains like medical imaging, obtaining such large datasets can be challenging and expensive \cite{diaz2024monai}. Therefore, our primary goal is to design a system that can produce accurate labels with a small amount of training data. {\tool} addresses this by using ILP to infer labeling rules, significantly reducing the initial data requirements compared to deep learning approaches.

\textit{DG2. \textbf{Supporting Iterative Refinement}} 
Automated methods are prone to errors, particularly in specialized domains \cite{10.1145/3655755.3655760, zhang2023peanut, 10.1007/978-3-031-21707-4_26, gygli2020efficient, 10.1145/1821748.1821847, li2024labelaid}. Human intervention is crucial for error reduction. Semi-automated methods offer a promising approach by facilitating iterative refinement, effectively combining the efficiency of automation with the expertise of human annotators. However, previous works exhibit two key limitations: First, primitive refinement support offers iterative refinement capabilities, but previous works lack innovative techniques for more effective support \cite{10.1145/3655755.3655760, 10.1145/3511176.3511207}. Second, previous approaches enable large-scale, efficient image labeling but fall short in effectively eliciting domain knowledge \cite{10.1145/3658271.3658329}. {\tool} addresses these challenges by (1) offering recommendations for refinement, including rule refinement based on holdout data accuracy, object recommendations for predicate refinement, and image suggestions for uncertain cases, and (2) enabling users to create and edit rules directly, allowing them to apply their domain knowledge to improve the automated labeling system.

\textit{DG3. \textbf{Enhancing Interpretability.}} The interpretability of AI-driven decisions is crucial, particularly in collaborative settings where human annotators must understand and trust the system's outputs \cite{amershi2019guidelines, 10.1145/3411764.3445315, 10.1145/3613904.3642780}. {\tool} addresses this critical need by enhancing the transparency and interpretability of its decision-making process. It achieves this goal through two key features. The first feature is Rule Visualization, which represents each rule in Domain-Specific Language (Boolean algebra) to provide clear insights into the system's logic; The second is a Visual Programming Interface, offering an intuitive way for users to directly interpret and modify rules.

\textit{DG4. \textbf{Reducing Use Barriers.}}  Image labeling is an example of an end-user programming task, where domain experts with specialized knowledge are required to annotate data but may lack programming skills. As identified by \citet{10.1109/VLHCC.2004.47}, a common barrier in end-user programming environments is the use barrier. It occurs when a user understands the purpose of a programming tool or feature but struggles to apply it correctly or effectively in practice. The use barrier is often rooted in limited knowledge of programming concepts, unfamiliarity with syntax, or difficulty in troubleshooting errors. Domain-specific data annotation often requires input from experts who possess specialized knowledge but may lack programming skills. Annotation tools that heavily rely on coding can impede these experts' ability to effectively elicit their domain knowledge. Visual programming systems have demonstrated that non-programmers can create complex programs with minimal training \cite{10.1145/22627.22349}. For instance, Scratch introduced a simplified visual interface for young users creating digital applications \cite{10.1145/1868358.1868363}. Drawing inspiration from this work, {\tool} designs visual programming components to minimize the learning curve. The visual programming component incorporates rule creation and editing with a simple and intuitive drag-and-play interface, enabling users to quickly elicit the domain knowledge and become proficient in using the system.

\section{System Design and Implementation}
In this section, we discuss the design and implementation of {\tool}. As mentioned earlier, {\tool} achieves image labeling by automated labeling rule generation, enabling manual rule editing, and supporting users with three ways of recommendations. \revise{We first briefly explain the system architecture and an overview of the user interface, then we explain how the system is initialized and each component in more detail.}

\subsection{\revise{System Architecture}}
\revise{{\tool} comprises three components: the pre-trained visual predicate extractor, the backend algorithm, and the interactive refinement user interface, as shown in Figure \ref{fig:sys_arc}.
For each image labeling task, users first define the predicates based on their domain knowledge and import corresponding pre-trained computer vision models for visual predicate extraction. 
Based on the user-defined predicates, the backend algorithm uses inductive logic programming to search for suitable image labeling rules and apply them to auto-label images (detailed in Section~\ref{sec:ilp}).
{\tool} proposes a novel user interface that visualizes the logic of the image labeling rules and presents the image labeling results.
Users can correct labels in a gallery view and directly edit rules through a visual programming interface (detailed in Section~\ref{sec:uio} and Section~\ref{sec:vpi}).
} 

\begin{figure*}[ht]
  \centering
  \includegraphics[width=.95\textwidth]{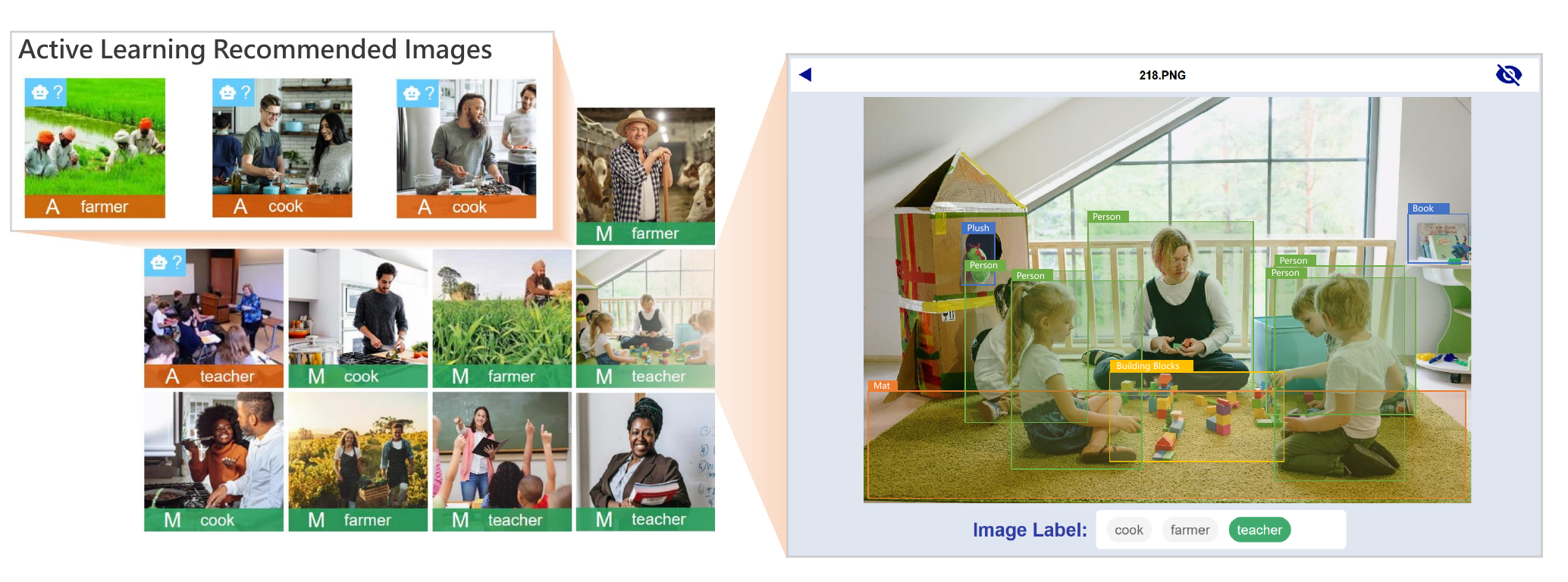}
  \caption{The image gallery of {\tool}. Users are allowed to view the extracted visual information when viewing images individually. Images that are recommended by active learning are highlighted and prioritized in the gallery.}
  \Description{Image Gallery Interface}
  \label{fig:gallery}
\end{figure*}

\subsection{User Interface Overview}
\label{sec:uio}
In each iteration of image labeling, users will interact with three parts of our system: the image gallery, the visualized labeling statistics, and the rule editor.

\textit{\textbf{Image Gallery: }} Users can explore and label images in the gallery, where each image is tagged by its labeling status (i.e., automatically labeled, manually labeled, or unlabeled). The gallery supports efficient browsing, filtering, and searching based on label and labeling status. Furthermore, as demonstrated in Figure \ref{fig:gallery}, {\tool} also supports users to view extracted information from images by pre-trained model to facilitate user labeling, such as detected objects with bounding boxes or segmentation masks. After the first iteration, several images will be prioritized in the gallery according to multi-criteria active learning, suggesting them for manual labeling (detailed in Section~\ref{sec:rec}).

\begin{figure}[htbp]
  \centering
  \includegraphics[width=\linewidth]{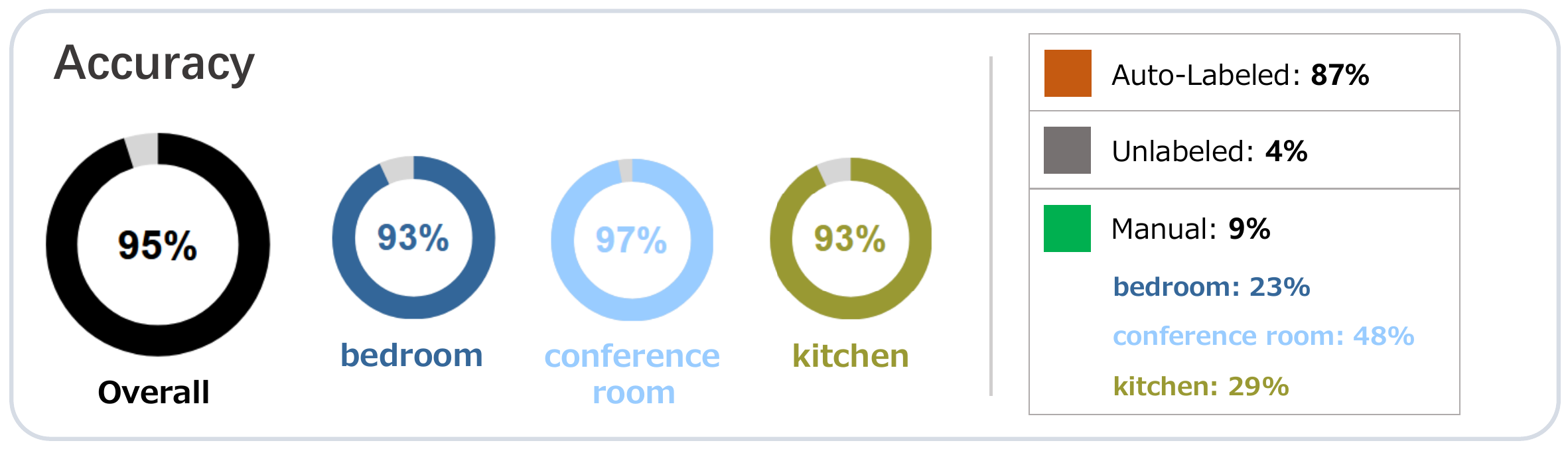}
  \caption{Rule accuracy and labeling progress statistics visualized by {\tool}.}
  \label{fig:label_sta}
\end{figure}

\textit{\textbf{Labeling Progress Statistics: }} In each iteration, real-time labeling progress statistics, such as the percentage of automatically labeled images, are visualized in the interface of {\tool} (Figure \ref{fig:label_sta}). This information enables users to adjust labeling strategies, such as prioritizing which rule to refine or which image to label. Rule performance statistics will be further detailed in Section~\ref{sec:rec}.

\begin{figure*}[htbp]
  \centering
  \includegraphics[width=\textwidth]{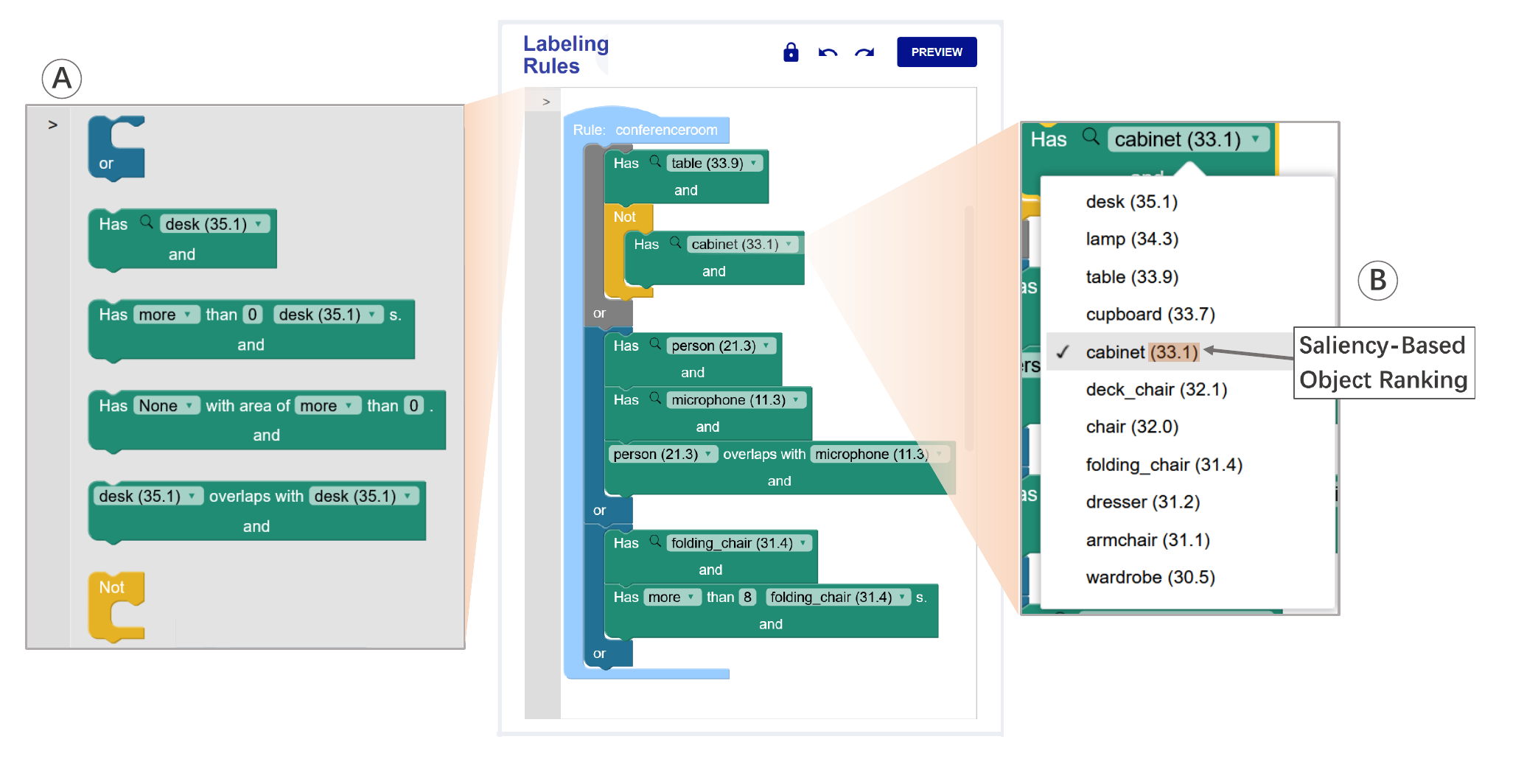}
  \caption{The interface of the rule editor in {\tool}. Users are able to add different types of predicates and clause operators via drag-and-drop (A). Users can select objects from the drop-down list, which they are ranked by a saliency-based method (B).}
  \Description{Rule Editor}
  \label{fig:rule_editor}
\end{figure*}

\textit{\textbf{Labeling Rule Editor: }} Users can refine labeling rules through a visual programming interface (Figure \ref{fig:rule_editor}). The operations supported by the rule editor are discussed in detail in Section~\ref{sec:vpi}.

\subsection{Labeling Rule Generation}
\label{sec:ilp}
We use the same ILP \cite{Muggleton_1994} algorithm as RAPID \cite{Wang_Tu_Xiang_Zhou_Chen_Li_Zhang_2023}, which is the FOIL algorithm \cite{Quinlan_1990}. It generates logic rules by iteratively selecting predicates that best differentiate between positive and negative examples. In {\tool}, FOIL inductively infers labeling rules for each class based on visual predicate extracted from human-labeled images, such as detected objects. These rules are expressed as a disjunction of clauses, each containing a conjunction of predicates. 

\begin{figure*}[htbp]
  \centering
  \includegraphics[width=0.7\linewidth]{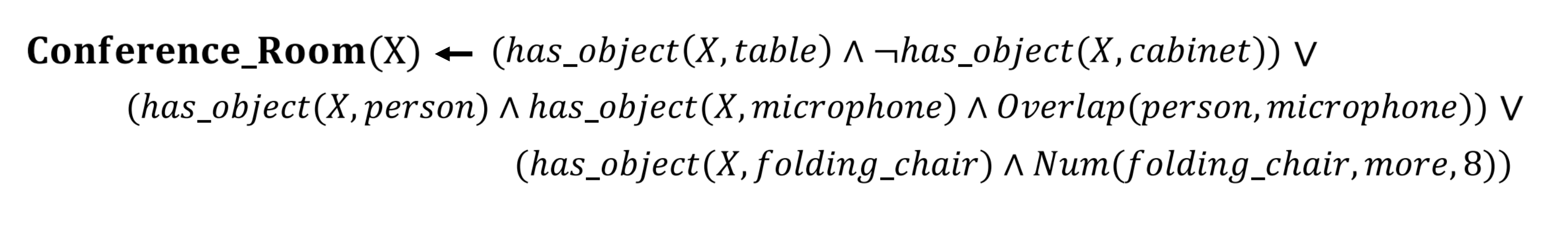} 
  \caption{Labeling Rule Example 1: Classify indoor scene images as a conference room.}
  \Description{Rule Format}
  \label{fig:rule_format_common}
\end{figure*}

{\tool} supports a variety of rule predicates, including but not limited to objects in the scene, the overlapping relationship between objects, and object appearance frequency. All predicates can also be negated in the expression. For instance, the rule in Figure \ref{fig:rule_format_common} identifies an indoor scene as a conference room. It consists of three conjunctive clauses connected by the ‘OR’ operator. Inside each clause, visual predicates, such as ‘image contains at least one table,’ are connected by the ‘AND’ operator.  This rule means that the image is classified as a ‘conference room’ scene if the image contains at least one table but no cabinet, if the person and microphone exist in the image and they overlap, or if there are more than 8 folding chairs in the image.

\begin{figure*}[htbp]
  \centering
  \includegraphics[width=0.75\textwidth]{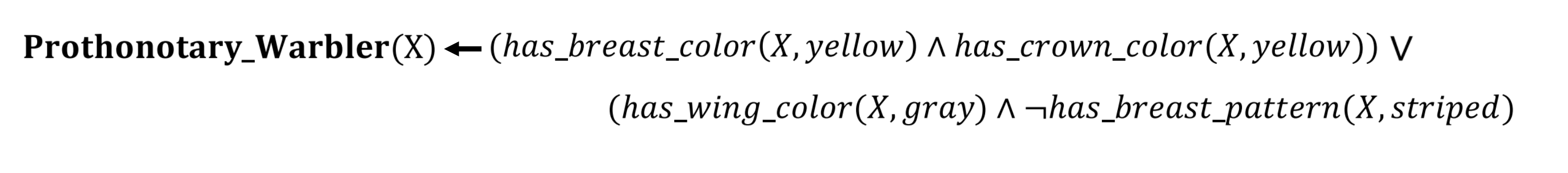} 
  \caption{Labeling Rule Example 2: Classify bird images as Prothonotary Warbler.}
  \Description{Rule Format}
  \label{fig:rule_format_bird}
\end{figure*}

Furthermore, {\tool} supports customized predicates for specialized image labeling tasks. For example, in Figure \ref{fig:rule_format_bird}, we use detected bird features as predicates in the rule of identifying the ‘Prothonotary Warbler.’ This rule means that the image is classified as a ``Prothonotary Warbler'' if the bird has both a yellow breast and yellow crown or if it has gray wings and no striped breast pattern. 

\subsection{Direct Rule Refinement via Visual Programming Interface}
\label{sec:vpi}
\begin{figure*}[htbp]
  \centering
  \includegraphics[width=0.8\textwidth]{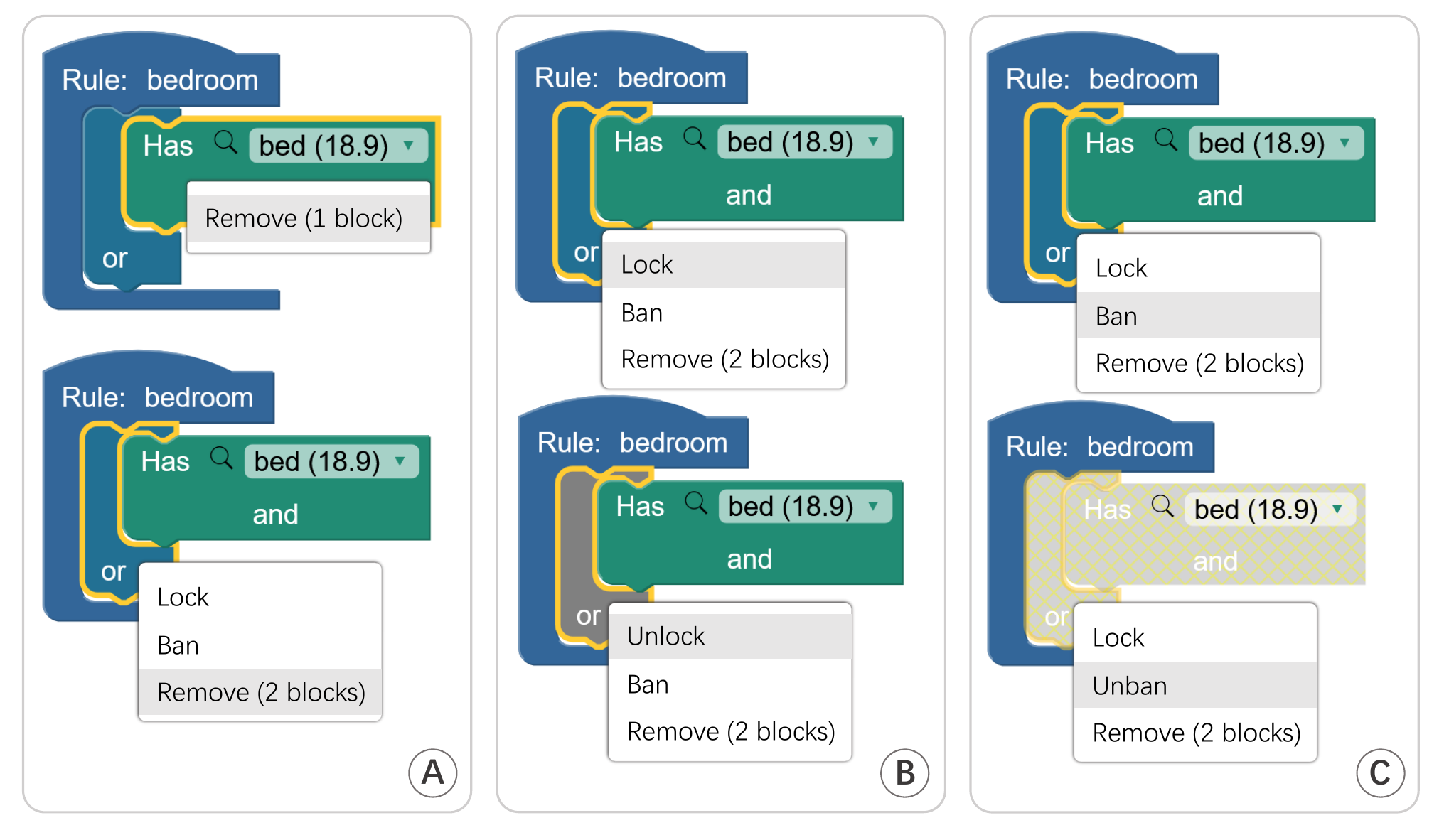}
  \caption{{\tool} provides different types of rule operations, including (A) removing rule predicates or clauses, (B) locking and unlocking rule clauses, and (C) banning and unbanning rule clauses. For locking and banning operations, the appearance of rules also changes accordingly.}
  \Description{Rule Operations}
  \label{fig:rule_operation}
\end{figure*}

{\tool} allows users to directly manipulate labeling rules in a visual programming interface (Figure \ref{fig:rule_editor}). This feature simplifies rule editing and refinement with the following operations:

\begin{enumerate}
 \item \textbf{[Drag and Drop]} operations allow users to easily restructure labeling rules. All rule components can be ``grabbed'' and dragged to a different location within the panel.
 \item \textbf{[Add]} a predefined rule predicate template, negation operator, or the ‘OR’ clause into labeling rule based on users’ intentions (Figure \ref{fig:rule_editor} A). Each clause must contain at least one predicate.
 \item \textbf{[Edit]} allows users to specify predicate value by entering or selecting from the drop-down list (Figure \ref{fig:rule_editor} B). Users can also search particular predicate values for quicker editing.
 \item \textbf{[Remove]} clauses and predicates from the labeling rule. In Figure \ref{fig:rule_operation} A, users can remove one ‘has bed’ block or the whole clause containing the ‘OR’ operator block and the ‘has bed’ block. This operation does not prevent removed predicates from being reintroduced in future rule generation.
 \item \textbf{[Lock]} operations enable users to preserve their manipulations by retaining the selected clause in the future generation. In Figure \ref{fig:rule_operation} B, the user locks the whole clause, and the background color of the ‘OR’ operator changes to gray, indicating a ‘locked’ status. Lock operations can be reversed by clicking ‘unlock,’ and the appearance of the ‘OR’ operator will change accordingly.
 \item \textbf{[Ban]} operations will exclude the selected clause from future generations. In Figure \ref{fig:rule_operation} C, the user bans the whole clause, and it changes to a grid background indicating a ‘banned’ status. Ban operations can be reversed by clicking ‘unban,’ and the appearance of the clause will change accordingly.
\end{enumerate}

\subsection{Recommendation Mechanisms}
\label{sec:rec}
\textit{\textbf{Refinement Based on Rule Performance Statistics.}} In machine learning, a holdout set is used to evaluate model performance on new data. Inspired by this approach, when each labeling task is created, a small set of labeled images, such as 10 images for each class, is also uploaded. The image set remains unseen during the labeling process. When labeling rules are generated, {\tool} calculates the accuracy of each rule on its holdout image set to help users identify rules that need more refinement. The performance panel (Figure \ref{fig:label_sta}) displays donut charts showing the accuracy of the overall holdout set as well as the accuracy for each label’s holdout set in the current iteration. Low accuracy in a particular class suggests that the corresponding labeling rule has a low labeling capability, indicating that further human refinement is required.

\textit{\textbf{Saliency-Based Rule Predicate Refinement Recommendations.}} When editing rule predicates, users may struggle to select specific objects from the many detected in the images. To address this, we use a saliency-based strategy that prioritizes significant objects. Inspired by the term frequency-inverse document frequency (TF-IDF) approach, {\tool} evaluates the importance of each object by adjusting for its frequency of appearance in the dataset. 

\begin{equation}
\text{Importance}_x = \frac{\sum_{i=1}^{N} \left( Objf_{x,i} \times \log\left(\frac{N}{imgf_x}\right) \right)}{N}
\end{equation} \\
Treating each image as a ``document'' and the entire dataset as a ``corpus,'' {\tool} calculates the importance of each object based on its average TF-IDF score (Equation 1) across all images when the dataset is uploaded. In Equation 1, $i$ denotes an image, $Objf_{x,i}$ denotes the frequency of object $x$ in image $i$,  $imgf_x$ denotes the number of images containing the object $x$, and $N$ denotes the total number of images. By this method, as shown in Figure \ref{fig:rule_editor} B, objects with low importance scores will be placed toward the bottom of the drop-down list, which helps users avoid selecting common objects that do not contribute significantly to the rule refinement. Furthermore, selecting prioritized objects could benefit image labeling. A high score indicates that an object is particularly relevant to a group of images, making it a strong candidate for use as a labeling criterion, especially for classes that depend on the consistent appearance of a specific object.

\textit{\textbf{Labeling Recommendations by Multi-criteria Active Learning.}} For users unfamiliar with visual programming interfaces, directly editing rules can be challenging. To simplify this process, {\tool} allows users to refine rules by providing additional image examples for learning and rule regeneration. {\tool} uses the multi-criteria active learning algorithm from RAPID \cite{Wang_Tu_Xiang_Zhou_Chen_Li_Zhang_2023} to suggest images for labeling based on their diversity and informativeness, using vectorized visual predicates rather than raw pixel data. 

For the diversity criterion, we employ a clustering algorithm to select images. Each image is represented by a feature vector, with dimensions corresponding to visual predicates extracted from the image. Using k-means clustering, we select centroids from these clusters as representative and diversified samples. They are chosen because they maximize the separation between clusters, ensuring that each selected image represents a distinct group with unique visual characteristics. 

For the informativeness criterion, images are selected based on the situation where ambiguous images can be labeled by multiple labeling rules. When an image is not labeled or can be labeled by multiple rules, we calculate its informative score based on the number of satisfied rules and the average percentage of predicates in each unsatisfied rule that also exists in the image. A high informative score indicates the image can be or is close to being labeled by multiple rules.

To combine both criteria, {\tool} ranked all images according to their informativeness. Then, {\tool} selects top $n$ images as an intermediate set and applies diversity criterion on them to find centroid images. The final images are visually highlighted as active learning suggestions in the gallery by prioritization and marking them with a blue bot icon (Figure \ref{fig:gallery}).

\subsection{\revise{Pre-trained Extraction Model Selection}}
\revise{Users need to provide their images to a pre-trained visual predicates extractor. The choice of extractor can be tailored to the specific needs of the task. For instance, the Segment Anything Model (SAM)~\cite{SAM} performs particularly well when labeling rules require information from segmented objects, such as size. In our user study, BEAL has demonstrated strong segmentation performance on medical images~\cite{BEAL}. For labels that rely on object detection within images, DINO-based models~\cite{dino} such as Grounding-DINO~\cite{ground_dino} may also be effective as we use it to set up the common-domain task experiment. Since the input pipeline of {\tool} is model-agnostic, users have the flexibility to apply any computer vision methods suitable for their designed predicates and follow their expertise.}

\definecolor{myblue}{rgb}{0.1, 0.2, 0.7}
\definecolor{mygreen}{rgb}{0.44,0.68,0.28}
\section{User Scenario}

Dr. Franklin, an ecologist specializing in forest ecosystems, has been invited by a team of AI researchers to assist in labeling 1,000 images representing four distinct forest biomes. These images will serve as the training set for a new computer vision model. The project is challenging due to subtle visual distinctions between biomes, where environments may look similar but differ by specific species. Additionally, the complexity within each biome leads to varied features, making it hard for existing CV models to perform effectively, while manually labeling 1,000 images remains time-consuming and costly. To overcome these issues, Franklin turns to {\tool}.

\begin{figure*}[htbp]
  \centering
  \includegraphics[width=0.9\textwidth]{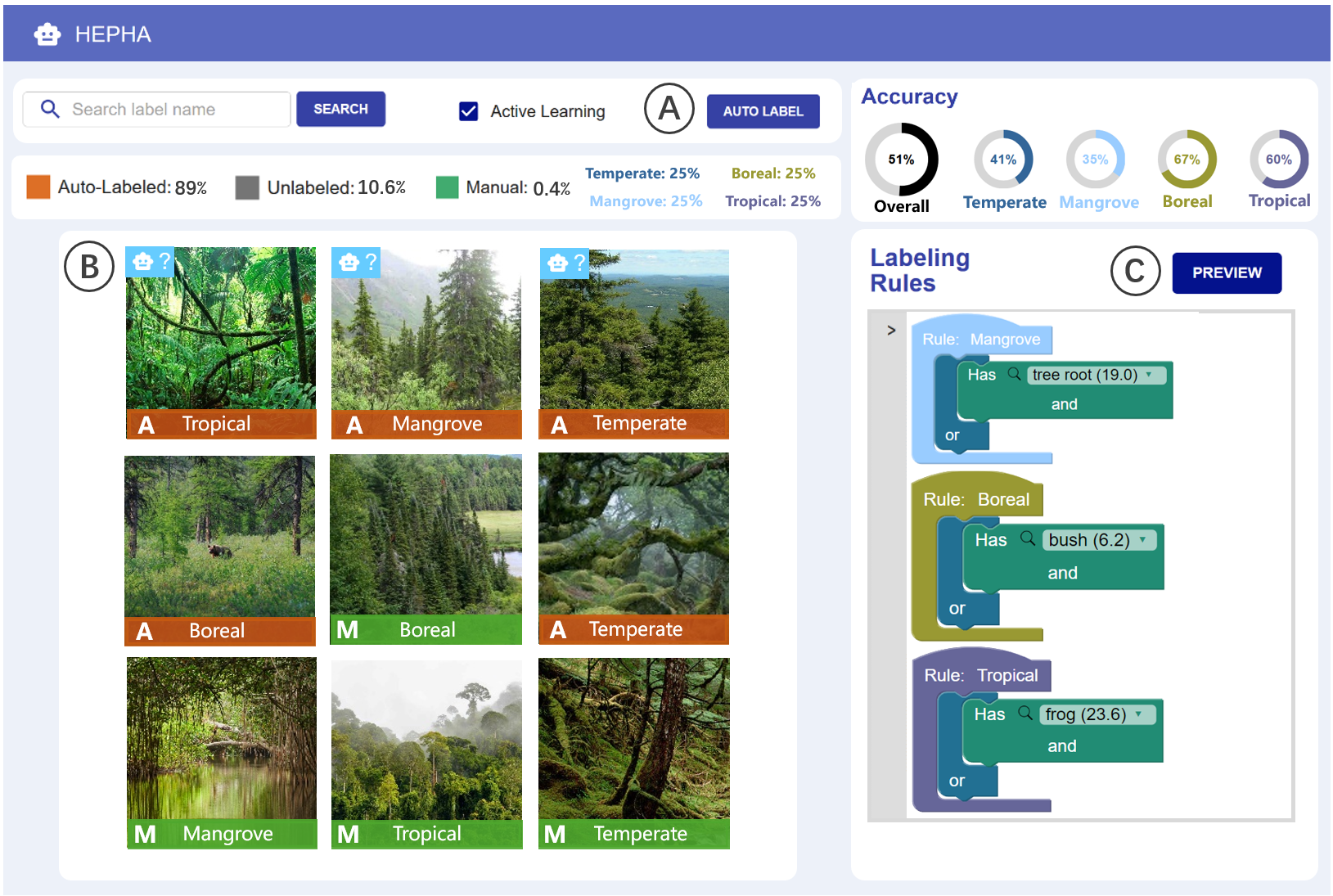}
  \caption{HEPHA Main Interface}
  \Description{User Scenario}
  \label{fig:user-scenario}
\end{figure*}

\textit{\textbf{Model Selection and Dataset Preparation}}. Dr. Franklin initiates the process by selecting a general-purpose object detection model, \textit{Grounding DINO}, to extract basic visual attributes from each image, such as trees, plants, and animals. He also prepares a test set of 80 images, representing roughly 10\% of the entire dataset, with 20 images for each forest biome, and loads both the raw images and the model’s feature outputs into {\tool} to facilitate the labeling process.

\textit{\textbf{Initial Labeling}}. Franklin starts by manually labeling a small subset of images. 
Upon inspecting one image from the gallery, Franklin identifies the presence of a large water pool with aerial roots, characteristics typical of a mangrove forest. 
He then assigns the label ``Mangrove'' to the image.
This procedure is repeated for one image per biome. Once the initial labels are assigned, Franklin activates the auto-labeling feature by selecting the \texttt{\textcolor{myblue}{\textbf{Auto Label}}} button (Figure \ref{fig:user-scenario}\textcircled{a}). This triggers the ILP rule generator, which derives logic rules for each biome class and applies them to the remaining images. For instance, based on the initial four labeled images, the model classifies images containing ``Tree Roots'' as Mangrove Forest and those with ``Bush'' as Boreal Forest. These rules are displayed in the Rule Section's visual programming interface, where Franklin can review and modify them as needed. Additionally, the Active Learning module highlights three ambiguous images for Franklin to review, allowing the rule generator to further refine the rules (Figure \ref{fig:user-scenario}\textcircled{b}). The initial labeling achieves an accuracy of 30\%, which is insufficient for model training. To improve accuracy, Franklin has two options: (1) label more images to capture distinguishing features or (2) refine the rules using his expertise.

\begin{figure*}[htbp]
  \centering
  \includegraphics[width=\textwidth]{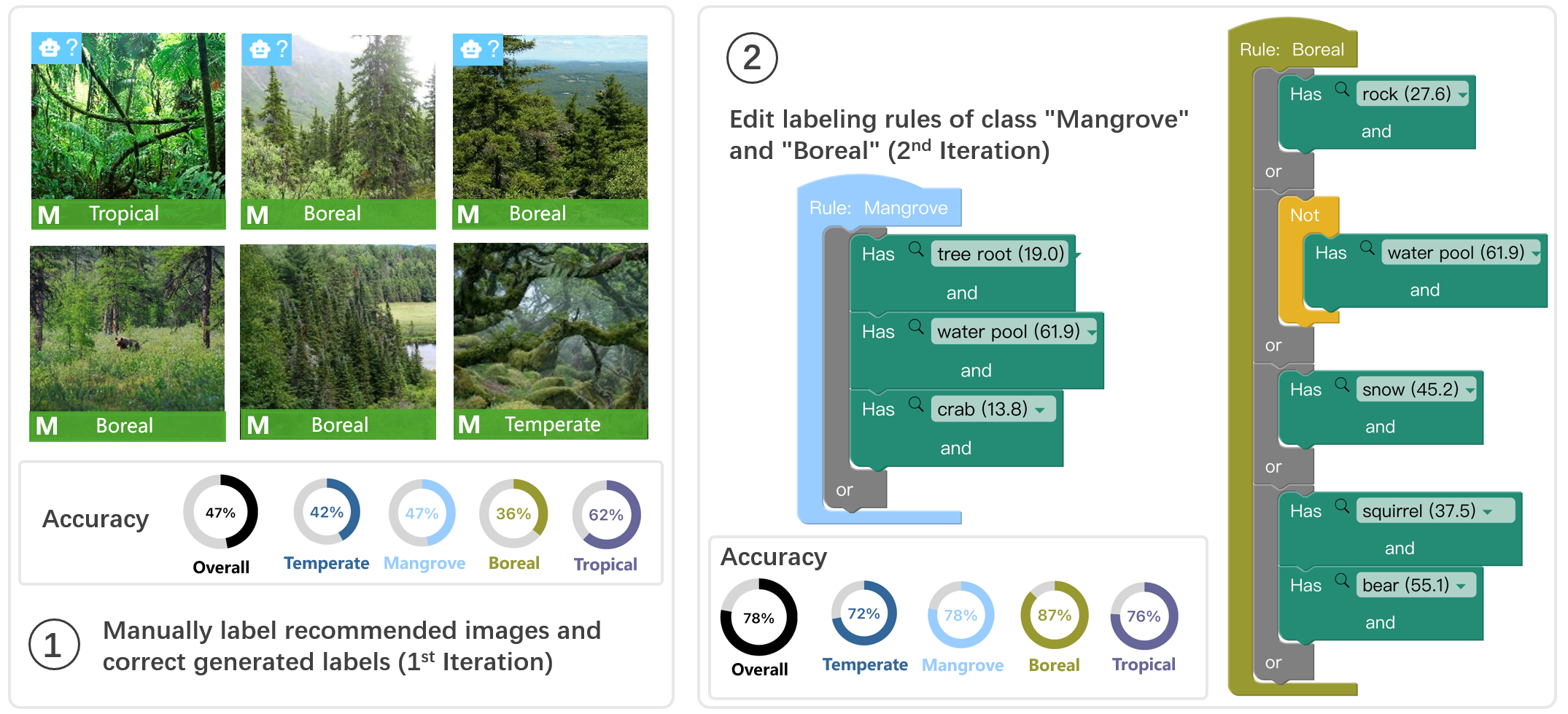}
  \caption{Iteration \textcircled{1}: Manual Labeling. The figure demonstrates the process of manually labeling recommended images and correcting generated labels in the first iteration. Iteration \textcircled{2}: Rule Refinement}
  \Description{User Scenario}
  \label{fig:us2}
\end{figure*}

    \textit{\textbf{Label with Active Learning}}. Given the limited number of initially labeled images, Franklin opts to label additional images. He begins with the images suggested by the active Learning module, correcting misclassifications, such as those incorrectly labeled as Mangrove. In total, Franklin labels 40 images and fixes some obvious misclassifications before running another auto-labeling task. The accuracy improves to 47\% (Figure \ref{fig:us2}\textcircled{1}).

\textit{\textbf{Rule Refinement}}. In the subsequent iteration, Franklin focuses on refining the generated rules, having identified that some predicates are overly broad or inaccurate. To strengthen the Mangrove rule, he adds more objects. He drags a \texttt{\textcolor{myblue}{\textbf{has-type}}} predicate from the rule block menu and selects relevant objects from the drop-down list. Unsure of which objects are most effective, he prioritizes objects with higher TF-IDF scores, indicating greater importance. He includes the predicates ``Has Water Pool'' and ``Has Crabs''---common attributes of a Mangrove forest. 
For the Boreal biome, he updates the rule ``Has Bush'' to ``Has Rock''.
Similarly, Franklin tweaks other rules based on his knowledge. To evaluate these modifications, he clicks the \texttt{\textcolor{myblue}{\textbf{Preview}}} button, applying the current rules to all images and recalculating accuracy (Figure \ref{fig:user-scenario}\textcircled{c}). Excited to see the overall accuracy jump from 47\% to 74\%, Franklin locks the modified predicates to ensure their continued use in future iterations. Despite this improvement, Franklin notices that the Boreal forest classification remains suboptimal, with an accuracy of 71\%. Upon further investigation, he revised the rule by adding the predicate ``Not Has Water Pool'' to exclude misclassified images with pools. Additionally, Boreal forest is the only biome where snowfall occurs during winter, leading Franklin to include a new clause with the predicate ``Has Snow''. This ensures that images satisfying any of these clauses will now be classified as Boreal Forest. Recalling that squirrels and bears are commonly found in Boreal environments, Franklin adds another clause to account for the presence of these animals. After implementing these adjustments, the updated rule increases the Boreal forest classification accuracy to 87\% (Figure \ref{fig:us2}\textcircled{2}.

Franklin continues refining rules and labeling the suggested images, with each iteration improving the model’s accuracy. After three more iterations, the overall labeling accuracy reaches 90\%. Satisfied with this result, Franklin finalizes the dataset, confident that the result is robust enough for training the CV model.

\section{User Study}

To understand how people interact with {\tool} in image labeling tasks, we conducted a within-subjects evaluation. The study was designed to compare {\tool} with its variant and a baseline deep learning approach across specialized and general domains.

\subsection{Baselines}

We implement two baseline methods to compare with {\tool}. First, ${\text{HEPHA}}_{NoEdit}$ is a variant of {\tool} without the rule-editing features. The users can only manually label images with assistance from active learning recommendations. This baseline falls under the category of semi-automated tools with active learning for refinement suggestion, similar to tools like Monai Label \cite{diaz2024monai}. Second, we build a fully automated tool by replacing the automated labeling algorithm with the ResNet18 model \cite{He_Zhang_Ren_Sun_2016}, a neural network for image classification tasks. ResNet18 is widely recognized for its strong performance, achieving 71.37\% accuracy in image classification tasks on the ImageNet dataset \cite{russakovsky2015imagenet}. Unlike {\tool}, this tool lacks the interpretability or refinement features: Users are only able to manually label, or correct images that have been misclassified by the system. It belongs to the category of fully automated tools that offer minimal interactive support \cite{acuna2018efficient, 10.1007/s11042-018-5973-x}.
 
\subsection{Tasks}

To evaluate \tool{}'s effectiveness in specialized domains, we constructed two datasets specifically for these areas. Each dataset allowed us to measure performance in more focused, domain-specific tasks. We also included two datasets from general domains to assess \tool{}'s generalizability across broader, more widely applicable contexts, ensuring that \tool{} performs well not only in specialized tasks but also in diverse, non-domain-specific scenarios. We designed four labeling tasks, with each task targeting one of these datasets. Specifically, the tasks involved the following four datasets:

\begin{itemize} 
    \item \textit{Bird}: A \textit{specialized}-domain dataset consisting of 400 images representing four easily-confused bird species: Bay-breasted Warbler, Cerulean Warbler, Magnolia Warbler, and Prothonotary Warbler. Each class has 100 images. Users are tasked to label the bird species for each class.
    \item \textit{Medical}: A \textit{specialized}-domain dataset comprising 116 images of glaucomatous eyes and 189 images of normal eyes. Each image includes both glaucoma diagnosis and structural segmentation. Users are tasked to label each eye as either normal or diseased.
    \item \textit{Occupation}: A \textit{general}-domain dataset containing 300 images of three occupations: chef, farmer, and teacher. Each occupation class has 100 labeled images. Users are tasked to label the occupation of the human in the image.
    \item \textit{Indoor Scene}: A \textit{general}-domain dataset consisting of 300 images depicting three types of indoor scenes: conference room, kitchen, and bedroom. Each scene category contains 100 labeled images. Users are tasked to label the class of each class.
\end{itemize}

\subsection{Protocol}

We recruited 16 participants through various academic channels, including university mailing lists and research group networks. The participants came from a diverse range of academic backgrounds, including 9 with majors in Computer Science or Data Science, 4 in Education, 2 in Statistics, and 1 in Psychology. In terms of programming experience, 4 participants had little to no experience, 2 participants had 1-2 years of experience, 6 had 3-5 years, and 4 had more than 5 years of experience. Regarding machine learning experience, 8 participants had little to no experience, 2 participants had less than 2 years of experience, and 6 participants had 3-4 years of experience. None of the participants had prior experience with image labeling. 

Each participant followed this procedure: 

\begin{enumerate} 
\item \textit{Pre-study survey}: Participants completed a pre-study survey designed to gather demographic information, technical background, prior experience with labeling tasks, and familiarity with image classification tools. This helped us understand the diversity of the participant pool and gauge their initial skill level.

\item \textit{Training session}: Each participant attended a 15-minute training session for all tools ({\tool} and the baseline tools). The session provided an overview of the user interface, features, and workflow. Participants are encouraged to ask questions and practice using each tool briefly to ensure they understand the labeling process and the tool’s capabilities before proceeding to the actual tasks.

\item \textit{Labeling sessions}: Participants are required to complete three labeling tasks covering both specialized and general domains. We randomly select three out of four datasets, ensuring that all datasets are assigned to an equal number of participants. During each task, participants were provided with a specific set of images to label within a 20-minute time frame. The order in which participants used the tools and completed the tasks was randomized to eliminate potential learning effects or biases resulting from the task sequence. Participants were instructed to complete each task within 20 minutes. They could stop labeling and finish the task early if they achieved over 60\% accuracy; otherwise, they were required to continue labeling until the end of the session. Throughout the task, participants were encouraged to verbalize their thought processes, explaining how they made labeling decisions, where they encountered difficulties, and which tool features were most helpful or limiting. After each task, participants completed a post-task survey, which captured their immediate feedback on the tool they just used. This included ratings on usability, perceived accuracy, confidence in their labels, and any frustrations or challenges they faced during the task.

\item \textit{Post-study survey}: At the end of the study, participants were asked to complete a comprehensive post-study survey. This survey collected their overall impressions of {\tool} and the baseline tools. Participants were asked to compare the tools and reflect on which tool they found most effective.
\end{enumerate}

The study was conducted remotely via Zoom, with participants accessing the labeling tools through a secure web interface. At the beginning of each session, we asked participants for their consent to record. Screen sharing was used to observe participant interactions and was recorded.

\section{User Study Result}
In this section, we report and analyze the difference in participants’ performance in the user study. For brevity, we denote the participants in the user study as P1-P16.

\subsection{User Performance}

\begin{figure*}[htp]
    \centering
    \includegraphics[width=0.8\textwidth]{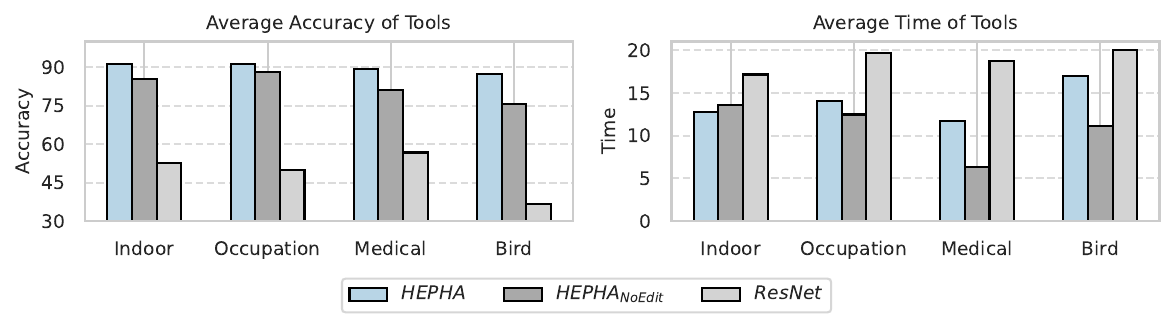}
    \caption{Accuracy and Time spent for different tasks}
    \Description{Result Section}
    \label{fig:acc-time}
\end{figure*}

\begin{figure}
        \centering
        \includegraphics[width=0.9\columnwidth]{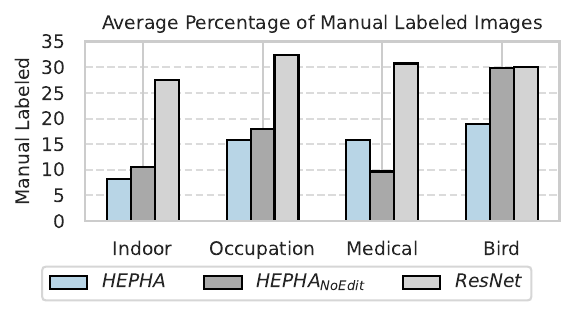}
        \caption{Proportion of labeled images to the whole dataset for different tasks}
        \label{fig:data_labeled}
\end{figure}

Figure \ref{fig:acc-time} presents the accuracy and time-spend of using three different tools---HEPHA, ${\text{HEPHA}}_{NoEdit}$, and ResNet---across four tasks in both general domain (Indoor, Occupation) and specialized domain (Medical, and Bird). Figure \ref{fig:data_labeled} presents the proportion of labeled images to the whole dataset when participants achieved the highest accuracy during the labeling process.

According to user study results, HEPHA consistently outperformed the other tools. It achieves the highest accuracy, over 87\%, in all 4 labeling tasks while requiring least amount of labeled images in 3 of them. Participants also spend less labeling time when using {\tool} compared to ResNet in both general-domain and specialized-domain tasks. In contrast, despite users labeled significantly more images using ResNet (Figure \ref{fig:data_labeled}), its accuracy was 41\% lower than HEPHA and 31.7\% lower than ${\text{HEPHA}}_{NoEdit}$ across all tasks.

Specifically, for specialized-domain tasks, {\tool} achieved over 87\% accuracy in both medical and bird datasets, which is 42\% higher than ResNet and 9\% higher than ${\text{HEPHA}}_{NoEdit}$ on average (Figure \ref{fig:acc-time}). In general-domain tasks, {\tool} attained over 91\% accuracy on both datasets, outperforming ResNet by 40\% and  ${\text{HEPHA}}_{NoEdit}$ by 9.6\%  on average (Figure \ref{fig:acc-time}). As {\tool} and ${\text{HEPHA}}_{NoEdit}$ both achieved significantly higher accuracy compared to ResNet, We can infer that applying ILP on extracted visual predicates to generate labeling rules is an effective image classification method, especially for similar images, such as eye images in the medical dataset. Furthermore, {\tool} outperformed ${\text{HEPHA}}_{NoEdit}$ by about 10\%, demonstrating that allowing users to edit rules is also significant to the labeling process. As one participant  (P11) commented in the final survey, ``\textit{I found that the closer HEPHA's labeling logic aligned with mine, the higher the accuracy.}'' 

We also measured participants' labeling time. For specialized-domain tasks, labeling with HEPHA took 3 to 7 minutes less than ResNet but 5 to 6 minutes more than ${\text{HEPHA}}_{NoEdit}$ (Figure \ref{fig:acc-time}). For general-domain tasks, users spent 5 to 6 minutes less time using HEPHA than using ResNet, with similar time spent ($\pm$1.5 minutes) compared to ${\text{HEPHA}}_{NoEdit}$. ResNet's significantly longer labeling time may be due to participants not achieving an ideal performance and utilizing the entire time limit during the task. In addition, the only way to improve ResNet's performance was by providing more images and correcting generated labels, which extended its training time. Furthermore, The slightly higher labeling time for HEPHA compared to ${\text{HEPHA}}_{NoEdit}$ could be attributed to two factors. First, users were more inclined to explore the rule-editing feature in {\tool}. As one participant (P3) noted, ``\textit{The rule editing feature is fun. I like to play around with it!}''  However, since rule editing features are disabled in ${\text{HEPHA}}_{NoEdit}$, fewer user operations may result in less labeling time. Second, since the participants were not experts in glaucoma diagnosis and bird detection, they needed additional time to apply tutorial knowledge to refine the rules effectively for specialized tasks.

{\tool} can also be evaluated by the number of images labeled during the study, calculated as the proportion of the total dataset labeled by participants.  In specialized-domain tasks, users labeled 13\% fewer images on average with {\tool} than with ResNet. Compared with ${\text{HEPHA}}_{NoEdit}$, users labeled 6\% more images in the medical dataset but 11\% fewer in the bird dataset using {\tool}. Overall, {\tool} and its variant required significantly fewer labeled images (16\%–19.5\%) to achieve the highest accuracy compared to ResNet, and 2.5\% fewer than ${\text{HEPHA}}_{NoEdit}$. In our study, since ResNet did not achieve satisfying accuracy, participants kept labeled images until the task ended, resulting in a much higher amount of labeled images. These results emphasize that using ILP-generated labeling rules in {\tool} achieves high accuracy with less training data compared to deep learning approaches. In addition, {\tool} achieved higher accuracy but less image labeling compared to ${\text{HEPHA}}_{NoEdit}$ indicating direct rule editing could be a more effective way to refine labeling rules than labeling additional images.

In summary, HEPHA not only outperforms other tools in terms of accuracy but also offers significant benefits in saving time and training images in both general-domain and specialized-domain tasks. Moreover, allowing users to edit labeling rules directly also contributes to more accurate labeling and fewer training data needs, reinforcing the effectiveness of human-AI collaboration in such settings.

\subsection{User Ratings of Individual Features}

\begin{figure*}[htbp]
    \centering
    \includegraphics[width=1\linewidth]{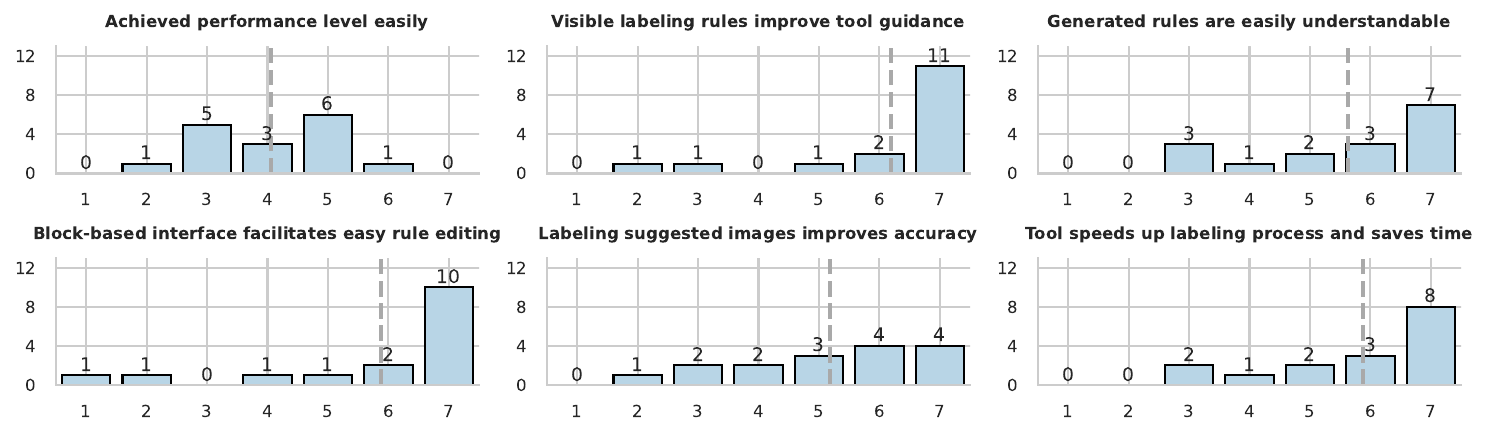 }
    \caption{User ratings of individual features}
    \label{fig:user_rating}
\end{figure*}

In the post-task survey, participants rated various features of HEPHA on a 7-point Likert scale (1—Strongly Disagree, 7—Strongly Agree). As shown in Figure \ref{fig:user_rating}, most participants responded positively to HEPHA's effectiveness in speeding up the labeling process and saving time, with 8 out of 16 participants strongly agreeing.

The rule editing features were also well-received by the majority of participants. The visualization of labeling rules was particularly appreciated, with 11 out of 16 participants strongly agreeing. 12 participants found the generated rules to be easily understandable. As P4 commented, ``\textit{The block-based design makes understanding the logic of the rules much easier. Each rule component is clearly connected within separate 'OR' blocks, making the structure and flow of the current rule straightforward to follow.}''. Furthermore, 10 out of 16 participants strongly agreed that the \textit{Block-based interface facilitates easy rule editing}. P5 noted,``\textit{Rule editing interface allows me to manipulate the system by myself, which increases the chance of getting labeling process right.}

Participants also responded favorably to the recommendation features, which were considered helpful for refining rules. All participants, including 10 who strongly agreed, acknowledged that viewing class accuracy enhanced the rule refinement process. For instance, in the final survey, P13 commented that ``\textit{I will check rule accuracy panel in each iteration to understand how and where should I improve the rules.}'' Furthermore, 10 out of 16 participants agreed that the predicate value suggestions were useful for refining the rules. P6 highlighted predicate value recommendation as a favorite feature. P6 said, ``\textit{Ranking more common objects at the bottom of the list helps me focus on more important features of the image.}'' Regarding the active learning image suggestions, 11 participants agreed that this feature helped improve the accuracy of the labeling rules. As P2 remarked, \textit{“I mostly labeled the images suggested by the system. The improvement of performance is good and steady.”}

\subsection{Cognitive Overhead}
\begin{figure*}[htbp]
    \centering
    \includegraphics[width=0.6\linewidth]{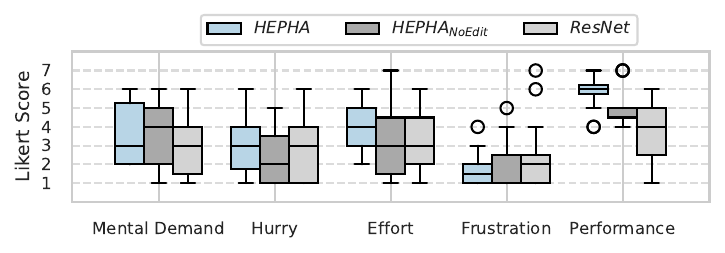}
    \caption{NASA Task Load Index Ratings in User Study}
    \label{fig:cognitive_load}
\end{figure*}

Figure \ref{fig:cognitive_load} presents participants' ratings across five cognitive factors from the NASA TLX questionnaire. There are no significant differences in mental demand, hurry, and frustration were found when comparing HEPHA to ${\text{HEPHA}}_{NoEdit}$ (Wilcoxon signed-rank test: $p = 0.7248$, $p = 0.4697$, $p = 0.7559$), or ResNet (Wilcoxon signed-rank test: $p = 0.1880$, $p = 0.7496$, $p = 0.2216$). These results indicate that more interaction, especially rule editing, do not introduce additional burdens or learning costs to users in the labeling process.

Regarding effort, participants using HEPHA reported significantly lower effort compared to ${\text{HEPHA}}_{NoEdit}$ (Wilcoxon signed-rank test: $p = 0.0422$), while differences between HEPHA and ResNet (Wilcoxon signed-rank test: $p = 0.2078$) were not statistically significant. Lower effort when using HEPHA means rule editing is an easier and more effective method for accuracy improvement. As P7 noted after using HEPHA, ``\textit{The editing process felt intuitive and effective. Whenever I had an idea to modify the rules, I could quickly check it by previewing the labeling result, which made the labeling process much easier and more efficient.}''

Furthermore, participants perceived significantly better performance with HEPHA compared to the ResNet ($p = 0.0009$), with a marginally significant difference when compared to ${\text{HEPHA}}_{NoEdit}$ ($p = 0.0613$). These results suggest that users found HEPHA more efficient and effective, particularly in comparison to less interactive tools like the ResNet.

\subsection{User Preference and Feedback}

\begin{figure}
        \centering
        \includegraphics[width=0.4\linewidth]{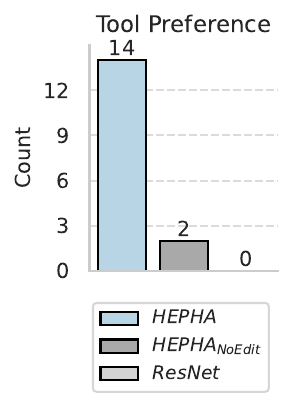}
        \caption{Tool preference from Post-study Survey}
        \label{fig:tool-preference}
\end{figure}

\begin{figure}
    \centering
    \includegraphics[width=\linewidth]{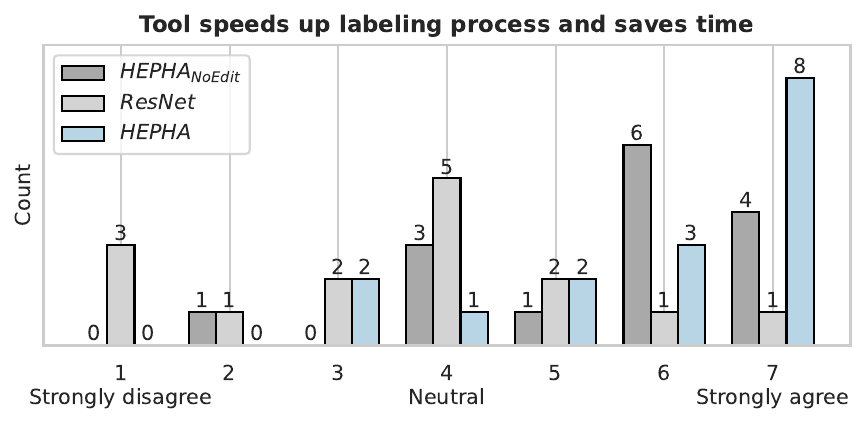}
    \caption{Comparison of users' perceived efficiency of each tool from the post-study survey.}
    \label{fig:score-comparison}
\end{figure}

In the post-study survey, participants self-reported their preferences among the three tools (as shown in Figure \ref{fig:tool-preference}) and perceived efficiency of each tool (as shown in Figure \ref{fig:score-comparison}).

As shown in Figure \ref{fig:tool-preference}, the majority of participants (14 out of 16) found HEPHA to be the most helpful approach. After coding the participants' responses, we identified two different themes that led to such a preference. First, 9 out of 14 participants endorsed rule editing features. For instance, P9 indicated that ``\textit{The interface allows me to understand and manipulate rules, so I could have some control on the model behavior. Also, compared to other baselines, HEPHA allows me to utilize my own knowledge, which helps to increase labeling accuracy.} 

Second, 7 out of 13 participants also appreciated the rich recommendations in each iteration as it assisted them during the labeling process. P13's feedback shows participants are exploring and mastering the recommendation feature to achieve high accuracy. P13 said, ``\textit{When I tried to edit labeling rules, I always refer to the accuracy. One trick I found is the rule with high accuracy might be overgeneralized.}'' P13 further explained, ``\textit{If we search on that label, you can see lots of images are mistakenly labeled to this class. If you correct them and regenerate the rules, accuracy usually significantly improves.}''

Furthermore, as shown in Figure \ref{fig:score-comparison}, half of the users strongly agreed with the statement that {\tool} speeds up the labeling process and saves time. In contrast, 4 out of 16 participants felt the same about ${\text{HEPHA}}_{NoEdit}$. This suggests that users consider the rule-editing features in HEPHA to have the potential to streamline the labeling process, demonstrating its greater time-saving advantage compared to image labeling during rule refinement.

Participants also mentioned some limitations of the system. Two participants suggested that the rule generation time could be faster, which is around 45 seconds on average per iteration. Another participant expressed the desire for more complex rule structures, such as allowing nested 'OR' clauses, highlighting a need for additional editing features. These insights will guide future exploration of the system to improve both performance and user experience. Limitations of current {\tool} will be detailed in Section 8.4.

\section{Discussion}
Deep learning approaches often require large amounts of training datasets, which can be challenging to acquire in specialized domains. According to our user study, utilizing the {\tool} system allows users to perform fast and accurate image labeling with little manual effort. The system’s interactive interface enables users to both understand the underlying model and integrate their domain expertise into the labeling process. In the following sections, we discuss the implications of our system design and user study findings, as well as the limitations and future directions for this research. 

\subsection {Eliciting Human Expertise in Automation Processes}
Traditional automation workflows often aim for full automation, relying on data-driven models that function with minimal human intervention. However, such systems are typically trained on large amounts of high-quality labeled data, which still requires large human efforts. In contrast, by eliciting human expertise, {\tool} achieves effective automated image annotation with reduced manual inputs, bridging the gap between manual and fully automated processes.

The ILP algorithm in {\tool} requires only a few images to generate labeling rules. Furthermore, the system allows users to correct image labels and \revise{directly edit the labeling rules. According to our user study results, the structured feedback from annotators leads to more efficient human corrections and enables {\tool} to achieve strong labeling performance with fewer human annotations.}
As P10 noted in the final survey, ``\textit{By allowing me to edit labeling rules, I can quickly create a bot that aligns with my thoughts and complete the remaining labeling on behalf of me.}''

\revise{While {\tool} is designed for image labeling tasks, our approach can be extended to other modalities, such as text labeling problems. For this extension, several adaptations would be necessary. First, we need to incorporate a pre-processing pipeline for text data to support rule induction, such as removing stop words and generating semantic embeddings as text features. Second, new types of rule predicates would need to be designed to capture linguistic structures and semantic information, including key phrases, syntactic patterns, and entity relationships, as opposed to the visual features used in image tasks. Third, users would refine these rules through a new specialized interface that is tailored for text-based tasks, which should support functionalities that can visually apply rules to each document and assist user validation. For example, changing the gallery (Figure \ref{fig:gallery}) interface to a text viewer.}

\subsection{Enhancing Transparency in ML System}
In our user studies, we found that a lack of interpretability in model outputs often leads to user frustration and hampers the refinement of models. For instance, after using the ResNet model on the Medical dataset, P6 expressed dissatisfaction, saying, \textit{“I have no idea how to proceed since there is little explanation about how the images are labeled.”} This highlights a critical issue in many ML systems: users are left without insight into how decisions are made, making it difficult for them to refine the system effectively.

{\tool} addresses this challenge by offering a transparent image labeling process where users can understand, debug, and refine the system's labeling logic. In each iteration, before optimizing the rules, detected objects and masks within images are visualized to help users review visual information and formulate their labeling logic, empowering them to guide {\tool} accordingly. As P5 noted, \textit{“When I’m unsure how to edit the rules, I check which objects are included in the images. Then, I change the rules and expect accuracy to improve.”} These insights enable users to accurately and adaptively apply their expertise to the data.

Furthermore, the underlying labeling logic is presented by the visual programming interface. The structure and content of each labeling rule are explained to support simple rule manipulations and intuitive rule operations. P3 commented, \textit{“The labeling rules were very easy to understand and intuitive to edit, especially with the ‘lock’ and ‘ban’ features.”} At this level, users can understand how the system makes decisions and provide targeted refinements. Moreover, throughout the labeling process, {\tool}'s interface provides continuous, real-time feedback, such as statistics on the proportion of manually labeled images and the effectiveness of specific rule adjustments. As P8 stated, \textit{“I keep receiving information from the interface. The statistics are always synchronized as long as I do corresponding operations.”} This ongoing flow of information ensures that users remain informed about the system's performance and can track the impact of their actions and refinements. 

For tools handling domain-specific tasks, experts need to be confident in the system's decisions and have the opportunity to improve it. By transforming the traditionally opaque machine-learning pipeline into a transparent and interactive system, {\tool} not only reduces the black-box effect but also enables users to play an active role in refining and debugging the model. A high transparency system interface can also be applied to the previous example of specialized document classification tasks. Extracted semantic information and phrases can be highlighted in the view of each document so that users can correspond to their expertise and correct generated labels. Furthermore, in text-related tasks, labeling rules and predicates could be further translated into human languages, even close to domain knowledge. This method could further enhance the overall interpretability of the labeling system and allow users to refine rules efficiently. 

\subsection{Automation vs. User Manipulation}

Many systems can be improved through either user interaction or automation. In {\tool}, users can manually edit labeling rules or label more images to refine and regenerate rules. In fact, if there is a strong alignment between user expertise and the task, it is possible to complete the task by mainly relying on rule editing. However, in our user study, participants often used a mixed approach, combining manual edits with image labeling. After the first iteration, they began with rule editing. When accuracy hit a bottleneck, they locked edited clauses and started labeling images. As ILP learned from more examples, new clauses were generated to enhance rule generalizability. P5 explained, \textit{“I love to lock up some very good rule components and label the suggested images to get more components from the system. A combination of generating and editing usually yields the best result.”} In this way, users can benefit from automatic rule refinement while avoiding complex manual manipulations.

According to our user study results, the majority of participants preferred using {\tool} for labeling images. However, when compared with ${\text{{\tool}}}_{NoEdit}$, users spent more time completing the labeling task with {\tool} due to increased interactions with the system, particularly in editing rules. Though both ${\text{{\tool}}}_{NoEdit}$ and the ResNet model are more automated approaches that require less human effort, users did not lose patience and interest in the system because of the need for more operations. On the contrary, increasing user involvement and control in the system may enhance user confidence and lead to better collaboration in completing tasks. However, high reliance on user operations and refinements could lead to frustration, higher labor costs, and longer task completion times. Therefore, an interactive system can be enhanced by achieving a better balance between automation and user manipulation.

\subsection{Limitations and Future Work}

\textit{\textbf{Balancing Rule Complexity and System Interpretability.}} In the user study, P4 stated that, ``\textit{I feel the rules are a bit simple. I wish it could have advanced features, such as nested logic. For example, if A is true, then either B or C must also be true, but not both.}'' Indeed, more complex rule structures could enable {\tool} to handle more challenging data labeling tasks. However, such complexity may also make the rules harder for users to interpret and manipulate, potentially hindering their ability to complete labeling tasks. Therefore, we initially developed {\tool} with a simpler disjunctive normal form (DNF) structure. In the future, we plan to explore more advanced rule structures to enhance {\tool}’s flexibility, enabling it to accommodate a wider range of domain expertise and meet the needs of expert users.

\revise{\textit{\textbf{Limitations of The Current Algorithm}} In the current implementation, {\tool} uses first-order logic as the rule format and FOIL for rule generation, prioritizing rule interpretability and computational efficiency. 
However, this design may encounter challenges when dealing with more complex domain-specific tasks and classification problems involving a large number of classes. 
For example, in medical image classification, images often contain a wide range of clinical features, such as cellular and glandular morphology \cite{cai2019human}. 
Labeling such images requires more precise visual predicates and more expressive rule structures.} In future work, we will explore alternative inductive logic programming (ILP) methods based on higher-order logic, such as $\lambda$PROGOL \cite{LamPROGOL}. It could generate more robust and more generalizable rules in complex structures, which may further enhance {\tool}’s performance in specialized domains.

Furthermore, user feedback highlights the need for more insightful design features. For instance, Participant P1 remarked, ``\textit{I would like to understand why a particular image cannot be labeled by the rule, so I could identify which predicates are missing.}'' To address this, future work should also focus on incorporating interactive features that provide explanations for rule application, enabling users to better understand and refine rules effectively.

\revise{\textit{\textbf{Limitations of User Study Design.}} In our current study in specialized domains, due to the challenges of recruiting domain experts for our experiment, our participants do not possess the same level of knowledge as real domain experts, such as ornithologists.}

\revise{In addition, 11 out of 16 participants had varying levels of coding background, which may have provided them with familiarity with concepts such as logic or machine learning. As a tool designed for domain expertise, it is code-free. The design of the labeling rules is also intentionally kept simple, relying solely on logical connectors such as ‘AND' and ‘OR’. In the future, we will further evaluate {\tool} on users without background in logical knowledge, and explore more intuitive and easy-to-understand logic visualization and representation designs.
}

\revise{Furthermore, the bird classification and glaucoma diagnosis tasks in our user study are relatively simple compared to more specialized tasks, such as those involving pathology or microscopy images. Therefore, the findings of this user study may not generalize to more challenging image labeling tasks, which should be addressed in future research.}

\section{Conclusion}

We present {\tool}, a mixed-initiative system that automates image labeling by generating and applying labeling rules through Inductive Logic Programming (ILP). {\tool} integrates a visual programming interface to enhance interpretability, enabling users to edit labeling rules based on their domain expertise. In addition, the system offers three recommendation mechanisms to guide and assist users in rule refinement. By enabling users to debug and take control of the automated labeling process, {\tool} allows them to leverage their knowledge to achieve accurate image labeling with minimal training data and human effort. To evaluate the usefulness and performance of {\tool}, we conducted a within-subjects user study with 16 participants on two specialized-domain datasets and two general-domain datasets. Our user study results demonstrated that users were able to achieve more accurate image labeling when using {\tool}. Furthermore, the labeling process was also faster comparing {\tool} with the deep learning method.

\bibliographystyle{ACM-Reference-Format}
\bibliography{sample-base, reference}

\end{document}